# Open problem in orthogonal polynomials

Abdulaziz D. Alhaidari

Saudi Center for Theoretical Physics, P.O. Box 32741, Jeddah 21438, Saudi Arabia
E-mail: haidari@sctp.org.sa
URL: http://www.sctp.org.sa/haidari

**Abstract**: Using an algebraic method for solving the wave equation in quantum mechanics, we encountered a new class of orthogonal polynomials on the real line. It consists of a four-parameter polynomial with continuous spectrum on the whole real line and two of its discrete versions; one with a finite spectrum and another with countably infinite spectrum. A second class of these new orthogonal polynomials appeared recently while solving a Heun-type equation. Based on these results and on our recent study of the solution space of an ordinary differential equation of the second kind with four singular points, we introduce a modification of the Askey scheme of hyper-geometric orthogonal polynomials. Up to now, these polynomials are defined by their three-term recursion relations and initial values. However, their other properties like the weight functions, generating functions, orthogonality, Rodrigues-type formulas, etc. are yet to be derived analytically. Due to the prime significance of these polynomials in physics and mathematics, we call upon experts in the field of orthogonal polynomials to study them, derive their properties and write them in closed form (e.g., in terms of hypergeometric functions).

## 1. Introduction

The wave function in quantum mechanics could be viewed as a vector field in an infinite dimensional space with local unit vectors. Therefore, in one of the formulations of quantum mechanics, the wave function at an energy $E$, $|\psi_E(x)\rangle$, is written as a bounded sum over a complete set of square integrable basis functions in configuration space with coordinate $x$:

$$|\psi_E(x)\rangle = \sum_n f_n(E) |\phi_n(x)\rangle, \qquad (1)$$

where $\{\phi_n(x)\}$ are the basis elements (local unit vectors) and $\{f_n(E)\}$ are proper expansion coefficients in the energy (the components of the wave function along the unit vectors). All physical information about the system, both structural and dynamical, are contained in these expansion coefficients. The *"Tridiagonal Representation Approach* (TRA)" is an algebraic method for solving the wave equation (e.g., the Schrödinger or Dirac equation) [1-4]. In the TRA, the basis elements are chosen such that the matrix representation of the wave operator is tridiagonal. Consequently, the resulting matrix wave equation becomes a three-term recursion relation for the expansion coefficients $\{f_n(E)\}$, which is solved in terms of orthogonal polynomials in some physical parameter(s) and/or the energy. If we write $f_n(E) = f_0(E) P_n(\varepsilon)$, where $\varepsilon$ is an appropriate function of the energy and physical parameters, then we have shown that $\{P_n(\varepsilon)\}$ is a complete set of orthogonal polynomials satisfying the said recursion relation with $P_0(\varepsilon) = 1$. The corresponding positive definite weight function is $[f_0(E)]^2$. These polynomials are associated with the continuum scattering states of the system where $E$ is a continuous set. On the other hand, the discrete bound states are associated with the discrete version of these polynomials.

We found all such polynomials that correspond to well-known physical systems and to new ones as well. For example, the scattering states of the Coulomb problem are associated with Meixner-Pollaczek polynomial whereas the bound states are associated with one of its discrete versions; the Meixner polynomial. Moreover, the scattering states of the Morse oscillator are associated with the continuous dual Hahn polynomial whereas the finite number of bound states are associated with its discrete version, the dual Hahn polynomial. Additionally, the continuum scattering states of the hyperbolic Pöschl-Teller potential correspond to the Wilson polynomial whereas the finite number of bound states are associated with the Racah polynomial, which is the discrete version of the Wilson polynomial. And so on.

Since 2005, however, we found a new class of exactly solvable problems that are associated with orthogonal polynomials, which were overlooked in the mathematics and physics literature [5-12]. These polynomials are defined, up to now, by their three-term recursion relations and initial value $P_0(\varepsilon) = 1$. However, their other important properties are yet to be derived analytically. These properties include the weight functions, generating functions, asymptotics, orthogonality relations, Rodrigues-type formulas, etc. Two classes of these new polynomials and their discrete versions are defined in the following two sections. In section 4, we introduce a modification to the hypergeometric polynomials in the Askey scheme [13,14] that manifest itself in a particular deformation of their corresponding three-term recursion relations.

## 2. The first polynomial class

The four-parameter orthogonal polynomial, which we designate as $H_n^{(\mu,\nu)}(z;\alpha,\theta)$, satisfies the following three-term recursion relation

$$(\cos\theta) H_n^{(\mu,\nu)}(z;\alpha,\theta) = \left\{ z\sin\theta \left[ \left(n + \tfrac{\mu+\nu+1}{2}\right)^2 - \alpha^2 \right] + \frac{\nu^2 - \mu^2}{(2n+\mu+\nu)(2n+\mu+\nu+2)} \right\} H_n^{(\mu,\nu)}(z;\alpha,\theta)$$
$$+ \frac{2(n+\mu)(n+\nu)}{(2n+\mu+\nu)(2n+\mu+\nu+1)} H_{n-1}^{(\mu,\nu)}(z;\alpha,\theta) + \frac{2(n+1)(n+\mu+\nu+1)}{(2n+\mu+\nu+1)(2n+\mu+\nu+2)} H_{n+1}^{(\mu,\nu)}(z;\alpha,\theta) \quad (2)$$

where $z \in \mathbb{R}$, $0 < \theta \leq \pi$ and $n = 1, 2, \ldots$. It is a polynomial of degree $n$ in $z$. Setting $z \equiv 0$ turns (2) into the recursion relation of the Jacobi polynomial $P_n^{(\mu,\nu)}(\cos\theta)$. Physical requirements dictate that $\mu$ and $\nu$ are greater than $-1$. The polynomial of the first kind satisfies this recursion relation together with $H_0^{(\mu,\nu)}(z;\alpha,\theta) = 1$ and

$$H_1^{(\mu,\nu)}(z;\alpha,\theta) = \frac{\mu-\nu}{2} + \frac{1}{2}(\mu+\nu+2)\left\{ \cos\theta - z\sin\theta \left[ \tfrac{1}{4}(\mu+\nu+1)^2 - \alpha^2 \right] \right\}, \quad (3)$$

which is obtained from (2) by setting $n = 0$ and $H_{-1}^{(\mu,\nu)}(z;\alpha,\theta) \equiv 0$. This polynomial has only a continuous spectrum over the whole real $z$ line. This could be verified numerically by looking at the distribution of its zeros for a very large degree. The asymptotics ($n \to \infty$) of $H_n^{(\mu,\nu)}(z;\alpha,\theta)$ could also be obtained numerically and found to be sinusoidal as a function of *n*, which is consistent with the expected physical behavior. Additionally, the physics of the problems associated with this polynomial suggests that it should have two discrete versions, one with an infinite spectrum (if $\alpha$ is pure imaginary) and another with a finite spectrum (for real $\alpha$). This is similar to the Meixner-Pollaczek polynomial and its discrete versions of the Meixner and Krawtchouk polynomials with infinite and finite spectra, respectively. The discrete spectrum is obtained as the set of values of *z*, which are pure imaginary (e.g., $\{-iz_k\}$), that make the amplitude of the sinusoidal asymptotics vanish. The size of this set is either finite or infinite. The two discrete versions of the polynomial are defined by their three-term recursion relations, which are obtained from Eq. (2) by the replacements $\theta \to i\theta$ and $z \to -iz_k$ giving

$$(1+\beta^2)\tilde{H}_n^{(\mu,\nu)}(k;\alpha,\beta) = \left\{ z_k(1-\beta^2)\left[\left(n+\tfrac{\mu+\nu+1}{2}\right)^2 - \alpha^2\right] + \tfrac{2(\nu^2-\mu^2)\beta}{(2n+\mu+\nu)(2n+\mu+\nu+2)} \right\} \tilde{H}_n^{(\mu,\nu)}(k;\alpha,\beta) \\ + \tfrac{4(n+\mu)(n+\nu)\beta}{(2n+\mu+\nu)(2n+\mu+\nu+1)} \tilde{H}_{n-1}^{(\mu,\nu)}(k;\alpha,\beta) + \tfrac{4(n+1)(n+\mu+\nu+1)\beta}{(2n+\mu+\nu+1)(2n+\mu+\nu+2)} \tilde{H}_{n+1}^{(\mu,\nu)}(k;\alpha,\beta) \tag{4}$$

where *k* is a non-negative integer of either finite or infinite range and $\beta = e^{-\theta}$ with $\theta > 0$. If we designate the discrete polynomials that solve (4) for the infinite or finite sets, $\{z_k\}_{k=0}^{\infty}$ or $\{z_k\}_{k=0}^{N}$, as $h_n^{(\mu,\nu)}(k;\alpha,\beta)$ and $g_n^{(\mu,\nu)}(k;N,\beta)$, respectively, then Table 1 shows some of the physical potential functions associated with the polynomials in this class. The size of the finite spectrum, *N*, is the largest integer less than or equal to $|\alpha| - \tfrac{\mu+\nu+1}{2}$.

## 3. The second polynomial class

While solving a Heun-type differential equation, we encountered recently another class of these new orthogonal polynomials [15]. It is also a four-parameter polynomial, which we designate as $Q_n^{(\mu,\nu)}(z;\alpha,\theta)$. It satisfies the following three-term recursion relation

$$(\cos\theta) Q_n^{(\mu,\nu)}(z;\alpha,\theta) = \left\{ z\sin\theta\left[\left(n+\tfrac{\mu+\nu+1}{2}\right)^2 - \alpha^2\right]^{-1} + \tfrac{\nu^2-\mu^2}{(2n+\mu+\nu)(2n+\mu+\nu+2)} \right\} Q_n^{(\mu,\nu)}(z;\alpha,\theta) \\ + \tfrac{2(n+\mu)(n+\nu)}{(2n+\mu+\nu)(2n+\mu+\nu+1)} Q_{n-1}^{(\mu,\nu)}(z;\alpha,\theta) + \tfrac{2(n+1)(n+\mu+\nu+1)}{(2n+\mu+\nu+1)(2n+\mu+\nu+2)} Q_{n+1}^{(\mu,\nu)}(z;\alpha,\theta) \tag{5}$$

where $z \in \mathbb{R}$, $0 < \theta \le \pi$ and $n = 1, 2, \ldots$. Note the inverse power on the square bracket, which constitutes a major difference from the recurrence relation (2). Here too, $Q_0^{(\mu,\nu)}(z;\alpha,\theta) = 1$ and $Q_1^{(\mu,\nu)}(z;\alpha,\theta)$ is obtained from (5) by setting $n = 0$ and $Q_{-1}^{(\mu,\nu)}(z;\alpha,\theta) \equiv 0$. This polynomial has a purely continuous spectrum over the entire real line. Moreover, it has another version, $G_n^{(\mu,\nu)}(z;\alpha,\beta)$, whose recursion relation is obtained from (5) by the replacement $\theta \to i\theta$ and $z \to -iz$ giving

$$\left(1+\beta^2\right)G_n^{(\mu,\nu)}(z;\alpha,\beta) = \left\{z\left(1-\beta^2\right)\left[\left(n+\tfrac{\mu+\nu+1}{2}\right)^2 - \alpha^2\right]^{-1} + \frac{2(\nu^2-\mu^2)\beta}{(2n+\mu+\nu)(2n+\mu+\nu+2)}\right\}G_n^{(\mu,\nu)}(z;\alpha,\beta)$$
$$+ \frac{4(n+\mu)(n+\nu)\beta}{(2n+\mu+\nu)(2n+\mu+\nu+1)}G_{n-1}^{(\mu,\nu)}(z;\alpha,\beta) + \frac{4(n+1)(n+\mu+\nu+1)\beta}{(2n+\mu+\nu+1)(2n+\mu+\nu+2)}G_{n+1}^{(\mu,\nu)}(z;\alpha,\beta) \quad (6)$$

where $z \geq 0$ and $\beta = e^{-\theta}$ with $\theta > 0$. If $\alpha$ is pure imaginary then the spectrum is purely continuous. However, if $\alpha$ is real then the spectrum is a mix of a continuous positive spectrum and a discrete negative spectrum of finite size $N$, where $N$ is the largest integer less than or equal to $|\alpha| - \tfrac{\mu+\nu+1}{2}$. In this case, the polynomial satisfies a generalized orthogonality relation of the form

$$\int_0^\infty \rho(z) G_n^{(\mu,\nu)}(z;\alpha,\beta) G_m^{(\mu,\nu)}(z;\alpha,\beta)\,dz + \sum_{k=0}^N \omega(k) G_n^{(\mu,\nu)}(z_k;\alpha,\beta) G_m^{(\mu,\nu)}(z_k;\alpha,\beta) = \lambda_n \delta_{n,m}. \quad (7)$$

where $\lambda_n > 0$, $\rho(z)$ and $\omega(k)$ are the positive definite continuous and discrete weight functions, respectively. The finite discrete spectrum $\{z_k\}_{k=0}^N$ could be determined from the condition that forces the asymptotics ($n \to \infty$) of $G_n^{(\mu,\nu)}(z;\alpha,\beta)$ to vanish.

## 4. Deformation of the Askey scheme of orthogonal polynomials

The results of our recent studies in [15] and [16], seem to suggest that the type of deformation in the recursion relation like that of the Jacobi polynomial in Eq. (2) is, in fact, common to a larger class of orthogonal polynomials: the Askey scheme of hypergeometric polynomials [13,14]. This scheme consists of two chains of hypergeometric orthogonal polynomials. One of them is a continuous set with the Wilson polynomial at the top of the chain that contains the continuous dual Hahn, continuous Hahn, Meixner-Pollaczek, Jacobi, Laguerre, etc. The other is a discrete set with the Racah polynomial at the top of the chain that includes, the dual Hahn, Hahn, Meixner, Krawtchouk, Charlier, etc. The polynomials in each chain are obtained from that at the top by certain limits of the hypergeometric functions (i.e., ${}_4F_3 \to {}_3F_2 \to {}_2F_1 \to {}_1F_1$). We write the three-term recursion relation of the original polynomials in the scheme generically as follows

$$xP_n^\gamma(x) = a_n^\gamma P_n^\gamma(x) + b_{n-1}^\gamma P_{n-1}^\gamma(x) + c_n^\gamma P_{n+1}^\gamma(x), \quad (8)$$

where $\gamma$ stands for a finite set of parameters and $x$ is a continuous or discrete set (finite or countably infinite) or both. As an example, for the Laguerre polynomial $L_n^\gamma(x)$, $x$ is continuous with $x \geq 0$, $\gamma > -1$ and $a_n^\gamma = 2n+\gamma+1$, $b_n^\gamma = -(n+\gamma+1)$, $c_n^\gamma = -(n+1)$. Now, the deformation of this recursion relation is introduced by modifying it such that it reads

$$x\tilde{P}_n^\gamma(x) = \left\{a_n^\gamma + \lambda\left[(n+\sigma)^2 - \alpha^2\right]\right\}\tilde{P}_n^\gamma(x) + b_{n-1}^\mu \tilde{P}_{n-1}^\gamma(x) + c_n^\mu \tilde{P}_{n+1}^\gamma(x), \quad (9)$$

where $\lambda$ is the deformation parameter and $\sigma$ is a function of the parameter set $\gamma$. As an example, the recursion relation (2) above is obtained by deforming that of the Jacobi polynomial $P_n^{(\mu,\nu)}(\cos\theta)$ where $x = \cos\theta$, $\lambda = z\sin\theta$ and $2\sigma = \mu+\nu+1$. Moreover, in Ref. [15] and [16],

–4–

we also encountered modified versions of orthogonal polynomials in the Askey scheme while searching for series solutions of the following second order linear differential equation

$$x(1-x)(r-x)\left[\frac{d^2 y(x)}{dx^2}+\left(\frac{a}{x}-\frac{b}{1-x}-\frac{c}{r-x}+d\right)\frac{dy(x)}{dx}\right]+\left(\frac{A}{x}-\frac{B}{1-x}-\frac{C}{r-x}+xD-E\right)y(x)=0, \quad (10)$$

where $\{a,b,c,d,r,A,B,C,D,E\}$ are real parameters with $r \neq 0,1$. For $d=0$, the equation has four regular singular points at $x=\{0,1,r,\infty\}$ and one of its solutions, which we referred to as "*generalized solution*" [15], is written as a series of square integrable basis functions like (1) with the expansion coefficients being modified version of the Wilson polynomial $\tilde{W}_n(z^2;\kappa,\tau,\eta,\xi)$ that satisfies the deformed recursion relation (9) where

$$x=z^2, \; \lambda=-r, \; 2\sigma=\kappa+\tau+\eta+\xi-1, \; \alpha^2=\tfrac{1}{4}(\kappa+\tau+\eta+\xi-1)^2-(\kappa+\eta)(\tau+\xi), \quad (11)$$

and the polynomial parameters $\{\kappa,\tau,\eta,\xi\}$ are related to the differential equation parameters in a particular way. In Ref. [17], the Authors refer to $\tilde{W}_n(z^2;\kappa,\tau,\eta,\xi)$ as the "*Racah-Heun polynomial*" but none of its analytic properties was given. On the other hand, for $d \neq 0$ Eq. (10) has four singularities with three regular at $x=\{0,1,r\}$ and one irregular at infinity. In Ref. [16], we obtained a series solution of this differential equation where the expansion coefficients are modified version of the continuous Hahn polynomial $\tilde{p}_n(z;\kappa,\tau,\eta,\xi)$ that satisfies the deformed recursion relation (9) with

$$x=\kappa+\mathrm{i}z, \; \lambda=\pm d^{-1}, \; 2\sigma=\kappa+\tau+\eta+\xi-1, \; \alpha^2=\tfrac{1}{4}(a+b+c-1)^2-D. \quad (12)$$

The polynomial parameters $\{\kappa,\tau,\eta,\xi\}$ are related to the differential equation parameters $\{a,b,c,d,r,A,B,C,D,E\}$ via one of two alternative ways depending on the $\pm$ sign of $\lambda$.

## 5. Conclusion

Due to the prime significance of these new (or modified) polynomials along with their discrete versions to the solution of various problems in physics and mathematics, we urge experts in the field of orthogonal polynomials to study them, derive their analytic properties and write them in closed form (e.g., in terms of hypergeometric functions). The sought-after properties of these polynomials include the weight functions, generating functions, asymptotics, orthogonality relations, Rodrigues-type formulas, Forward/Backward shift operator relations, zeros, etc.

# Table Caption:

**Table 1:** The physical potential functions associated with the new polynomial class of section 2. The polynomial parameters $\{\mu,\nu,\alpha,\theta\}$ are related to the potential parameters $\{V_0,V_1,V_\pm\}$ as shown, where $u_i = 2V_i/\eta^2$ and $\varepsilon = 2E/\eta^2$.

**Table 1**

| $V(x)$ | Polynomials | $\cos\theta$ | $\cosh\theta$ | $z$ | $\alpha$ | $\mu^2$ | $\nu^2$ | $\eta$ |
|---|---|---|---|---|---|---|---|---|
| $V_0 + \dfrac{V_+ - V_- \sin(\pi x/L)}{\cos^2(\pi x/L)} + V_1 \sin(\pi x/L)$ <br> $-L/2 \leq x \leq +L/2,\ (V_+ \pm V_-) \geq -2(\pi/4L)^2$ | $h_n^{(\mu,\nu)}(k;\alpha,\beta)$ | ----- | $\dfrac{\varepsilon_k}{u_1}$ | $\dfrac{1}{\sqrt{\varepsilon_k^2 - u_1^2}}$ | $u_0$ | $\tfrac{1}{4} + u_+ - u_-$ | $\tfrac{1}{4} + u_+ + u_-$ | $\pi/L$ |
| $\dfrac{1/4}{1-(x/L)^2}\left\{2V_0 + \dfrac{V_+}{(x/L)^2} + \dfrac{V_-}{1-(x/L)^2} + 4V_1\left[(x/L)^2 - \tfrac{1}{2}\right]\right\}$ <br> $0 \leq x \leq L,\ V_+ \geq -1/2L^2\ \ V_- \geq -2/L^2$ | $h_n^{(\mu,\nu)}(k;\alpha,\beta)$ | ----- | $\dfrac{\varepsilon_k - u_1}{\varepsilon_k + u_1}$ | $2\sqrt{-u_1\varepsilon_k}$ | $u_0 - u_1 - \tfrac{1}{16}$ | $1 + 2u_-$ | $\tfrac{1}{4} + 2u_+$ | $2\sqrt{2}/L$ |
| $\dfrac{1}{e^{\lambda x}-1}\left[V_0 + V_1(1 - 2e^{-\lambda x}) + \dfrac{V_+/2}{1 - e^{-\lambda x}}\right]$ <br> $x \geq 0,\ V_+ \geq -(\lambda/2)^2$ | $H_n^{(\mu,\nu)}(z;\alpha,\theta)$ <br> $g_n^{(\mu,\nu)}(k;N,\beta)$ | $-\dfrac{u_0}{u_1}$ | $-\dfrac{u_0}{u_1}$ | $\dfrac{1}{\sqrt{u_1^2 - u_0^2}}$ | $0$ | $-4\varepsilon$ | $1 + 2u_+$ | $\lambda$ |
| $\dfrac{V_+}{\sinh^2(\lambda x)} + 2\dfrac{V_0 + V_1[2\tanh^2(\lambda x) - 1]}{\cosh^2(\lambda x)}$ <br> $x \geq 0,\ V_+ \geq -\lambda^2/8$ | $H_n^{(\mu,\nu)}(z;\alpha,\theta)$ <br> $g_n^{(\mu,\nu)}(k;N,\beta)$ | $-\dfrac{u_0}{u_1}$ | $-\dfrac{u_0}{u_1}$ | $\dfrac{2}{\sqrt{u_1^2 - u_0^2}}$ | $-\dfrac{1}{16}$ | $-\varepsilon$ | $\tfrac{1}{4} + u_+$ | $\sqrt{2}\lambda$ |
| $\dfrac{V_0 + V_1 \tanh(\lambda x)}{\cosh^2(\lambda x)}$ <br> $-\infty < x < +\infty$ | $H_n^{(\mu,\nu)}(z;\alpha,\theta)$ <br> $g_n^{(\mu,\nu)}(k;N,\beta)$ | $-\dfrac{u_0}{u_1}$ | $-\dfrac{u_0}{u_1}$ | $\dfrac{1}{\sqrt{u_1^2 - u_0^2}}$ | $-\dfrac{1}{4}$ | $-\varepsilon$ | $-\varepsilon$ | $\lambda$ |